# Determining Causality in Travel Mode Choice

Rishabh Singh Chauhan[1,*], Christoffer Riis[2], Shishir Adhikari[3], Sybil Derrible[1], Elena Zheleva[3], Charisma F. Choudhury[4], Francisco Câmara Pereira[2]

[1]Department of Civil, Materials, and Environmental Engineering, University of Illinois Chicago, Chicago, 60607, United States

[2]DTU Management, Transport Division, Technical University of Denmark Kgs. Lyngby, Denmark 2800

[3]Department of Computer Science, University of Illinois Chicago, Chicago, 60607, United States

[4]Institute for Transport Studies School of Civil Engineering University of Leeds, Leeds, United Kingdom LS2 9JT

[*]Rchauh6@uic.edu

## Abstract

This article presents one of the pioneering studies on causal modeling in travel mode choice decision-making using causal discovery algorithms. These models are a major advancement from conventional correlation-based techniques. We propose a novel methodology that combines causal discovery with structural equation modeling (SEM). This modeling approach overcomes some of the limitations of SEM by combining the strengths of both causal discovery and SEM. Causal discovery algorithms determine causal graphs from observational data and domain knowledge, and SEMs estimate direct causal effects and test the performance of causal discovery algorithms. In this study, we test four causal discovery algorithms: Peter-Clark (PC), Fast Causal Inference (FCI), Fast Greedy Equivalence Search (FGES), and Direct Linear Non-Gaussian Acyclic Models (DirectLiNGAM). The results show that DirectLiNGAM based SEM model best captures causality in mode choice behavior. It passes several goodness-of-fit tests, including Root Mean Square Error of Approximation (RMSEA) and Goodness-of-Fit Index (GFI), and it achieves the lowest Bayesian Information Criterion (BIC) value. The analyses are conducted on data collected from the 2017 National Household Travel Survey in the New York Metropolitan area.

## 1. Introduction

Understanding and modeling travel mode choice behavior is a classic problem in transportation[1]. The long-established practice for conducting these studies is by statistical models based on utility maximization theory, such as multinomial logit models, and nested logit models[2–5]. These models make strong assumptions about the underlying relationships between the variables that are often considered unrealistic[3]. To overcome some of the limitations of conventional models, several machine learning-based methods for travel choice modeling have been proposed, including neural networks, extreme gradient boosting, random forest, and decision trees to name a few[2,3,6–8].

Despite their superiority, statistical and machine learning modeling techniques are still based on correlations[9] but as is well known, correlation does not imply causation. Two variables can have a mathematical relation (or correlation) between them without having a causal relation—a condition which is called 'spurious correlation'[10,11]. This is the reason behind the high correlation between extremely unlikely pairs of variables like cheese consumption versus fatal bedsheet tangling accidents and beef consumption versus deaths by lightning[12]. These examples highlight that correlations can be misleading and, thus, making policies based on correlations can be flawed.

Another limitation of commonly used travel mode choice models is that they assume that the explanatory variables affecting mode choice are independent of one another; i.e., these models do not consider the interaction between the factors influencing mode choice. This assumption is often unrealistic since the decision-making process is likely complex, with several variables directly or indirectly affecting mode choice decisions. For example, variables like household income and number of household members or age and education level are generally not independent to one another.

To address these issues, this article proposes a causality-based modeling approach for travel mode choice. Such models provide a complex graphical representation of the causal (or generative) mechanisms involved in mode choice decision-making. To estimate these causal models, we introduce the following novel procedure:

- **First, the causal links are determined using causal discovery algorithms:** Causal discovery is the process of extracting causal structures from observational and/or experimental data[13]. It is grounded in the complex concepts of causality and rooted in fields of statistics, economics, epidemiology, computer science, philosophy, and others[14]. Since causal discovery deals with causation, and not simply correlations, they are well suited to provide policy implications. Several causal discovery algorithms have been proposed in the literature, which differ in methodology and assumptions[15]. We selected four popular algorithms for this analysis: Peter-Clark (PC), Fast Causal Inference (FCI), Fast Greedy Equivalence Search (FGES), and Direct Linear Non-Gaussian Acyclic Models (DirectLiNGAM). These algorithms represent a diverse set of causal discovery algorithms where each follows a unique methodology, as explained in the methods section.
- **Second, Structural Equation Models (SEM) are used to (1) determine the most suitable causal model and (2) compute the quantitative measure of the direct causal effects between the variables:** SEMs have been previously used to model mode choice[16–19]. They possess some distinct advantages over most other techniques. In particular, SEMs: (i) are capable of handling exogenous, endogenous, and latent variables; (ii) can account for indirect, multiple, and reverse relationships; (iii) accept non-normal data; (iv) offer easier visualization of the modeled network; and (v) can potentially provide causal estimates[16,20]. However, despite these advantages, SEMs have their limitations as well. A key limitation is their inherent property to be confirmatory tools instead of exploratory. Therefore, the modeler needs to provide a hypothesized structural graph before creating the SEM[21]. As Bollen and Pearl (2013) pointed out, "…the SEM represents and relies upon the causal assumptions of the researcher. These assumptions derive from the research design, prior studies, scientific knowledge, logical arguments, temporal priorities, and other evidence that the researcher can marshal in support of them. The credibility of the SEM depends on the credibility of the causal assumptions in each application."[22] Given the importance of these causal assumptions, constructing a reasonable, hypothesized causal graph can be challenging. This might be particularly true in the case of large and complicated networks, like the mode choice decision-making process. Existing studies have usually had to rely only on domain knowledge and literature to build causal graphs. The lack of a dependable mechanism to generate a credible causal graph can create a substantial barrier in estimating a credible SEM. A standard, yet flawed, process to circumvent this limitation in SEM building is to alter a graph until the SEM achieves reasonable accuracy. This process is deemed controversial and can lead to an overfitted, unstable, unreliable, and incorrect model[23–26]. Therefore, we propose using causal discovery (from step 1) as a precursor to the SEM. The causal structure obtained from causal discovery can be fed as an input to a SEM model. This new step helps overcome the above-mentioned limitation of the SEMs.

Ultimately, this union between causal discovery and SEM is mutually beneficial. Causal discovery gains the quantitative measures of the causal effect from the SEM, while the SEM benefits from the data-driven causal graphs extracted by the causal discovery algorithms to be used as its input.

Overall, the objectives of this study are to:

i. Apply causal discovery algorithms to determine a graphical causal model of travel mode choice decision-making;
ii. Compare the performance of the various causal discovery algorithms to determine the most appropriate algorithm for studying travel mode choice;
iii. Estimate quantitative causal effects among the variables affecting the travel mode choice;
iv. Introduce a novel methodology that combines causal discovery and an SEM to model travel mode choice.

Causal modeling of travel mode choice has been rarely attempted. Xie and Waller[4] applied a Bayesian Network and used both observational information along with cause-effect hypotheses to learn a causal graph for mode choice prediction. Similarly, Ma et al.[5] used Structure Learning and unsupervised Bayesian Networks with domain knowledge to infer a causal graph. They explored three different classes of learning algorithms: constrained-based algorithms, score-based algorithms, and model averaging. More recently, Monteiro[27] used a constrained-based causal discovery algorithm, Find One-Factor Cluster (FOFC), to estimate a causal graph for travel satisfaction as well as mode choice. The inferred graph was compared with a graph constructed based on domain knowledge. It was found that FOFC recovered many of the cause-effect relations but had an undesirable property of being dependent on the order of the inputs. They also saw some potential in FOFC to contribute to the hypothesis generation for an SEM. Previously, studies like Golob and Hensher[18] have used an SEM to model mode choice and provided causal interpretations to their findings. These studies, however, had to rely on hypothesized causal graphs derived from domain knowledge and previous studies.

Our study seeks to fill the gap in the research in two important ways. First, it applies, analyzes, and compares four causal discovery algorithms that have never been used before for mode choice modeling. These algorithms can not only confirm well established causal hypotheses, but they can also reveal new causal relationships in the data. Thus, this study contributes an important analysis of the usage of causal discovery in transportation. Second, our proposed approach of combining causal discovery algorithms and SEMs can support modelers in making causal assumptions that are data driven, more substantiated, and more reliable than those usually made in the past studies.

## 2. Methods

### 2.1. Key concepts

This subsection explains the main concepts relevant to this study.

*Structural causal model (SCM):* A structural causal model[28] is used to model the causal assumptions in a domain by representing the relevant features and their interactions. A SCM, represented by *M(V, U, f)*, models how nature assigns values to the variables of interest using a set of variables *U*, *V*, and a set of functions *f* that assign values to each variable in *V* using other variables. The variable set *U* is termed as exogenous variables, which are external to the causal model and often considered errors or disturbances. The value of an endogenous variable $V_i \in V$ is explained by a function $f_i \in f$ of at least one exogenous variable $U_i \in U$ and optionally other endogenous variables; i.e., $V_i = f_i(V_{pa}, U_i)$, where $V_{pa} \subset V \setminus V_i$ is a set of direct causes of $V_i$. Therefore, a variable is a function of its known direct causes and unknown disturbances.

*Causal graphical model:* The causal relationship among the variables in an SCM can be represented using a directed acyclic graph (DAG) represented as *G(V, E)*, where *V* and *E* are a set of nodes (a.k.a. vertices) and links (a.k.a. edges) respectively. Each node (or vertex) has incoming links from its direct causes. $V_i \rightarrow V_j$ denotes a link from $V_i$ to $V_j$, where $V_i$ is the parent and $V_j$ the child. Two nodes are adjacent if there is a link between them. A directed path is a sequence of nodes obtained following the direction of the link. A graph is directed acyclic if there are no directed paths with repeated nodes. The nodes preceding the tail node of the directed path are the ancestors of the tail node. Similarly, the nodes following the head node of the directed path are descendants of the head node. Let the symbols *pa($V_i$, G)*, *anc($V_i$, G)*, and *des($V_i$, G)* indicate the sets of parents, ancestors, and descendants of $V_i$ for graph G respectively. In the Markovian case, the exogenous variables are assumed to be independent of one another and are not explicitly shown in the graph.

*Conditional independence relations:* The data generated by an SCM should adhere to the conditional independence relations that the causal graphical model entails. The joint probability distribution *P* described by *G(V, E)* factorizes to the product of the conditional probability of each random variable given its parents according to the causal Markov assumption, i.e.,

$$P(V_1, V_2, \ldots, V_n) = \prod_{i=1}^{n} P(V_i | pa(V_i, G)) \qquad (1)$$

The factorization in equation (1) follows the chain rule of probability theory and conditional independence (CI) relations entailed from the Causal Markov assumption. With the causal Markov assumption, a random variable is independent of all other variables except its parents and its descendants conditioned on the parents, i.e., $V_i \perp V \setminus \{V_i \cup pa(V_i, G) \cup des(V_i, G)\} \mid pa(V_i, G)$. The Markov conditions are not all the conditional independence relationships captured by the causal model. The notion of *d*-separation[29] is used to read off all the conditional independencies that hold for any data distribution that is generated by the mechanism described by a graphical model. The rules of *d*-separation are formally defined with the help of three sub-graph structures: (1) a chain, $V_i \rightarrow V_j \rightarrow V_k$, with an unidirectional path, (2) a fork, $V_i \leftarrow V_j \rightarrow V_k$, with a common cause, and (3) a collider, $V_i \rightarrow V_j \leftarrow V_k$, with a common effect. An undirected path is said to be blocked by a node $V_j$ with a conditioning set S of observed variables if one of two conditions hold: (i) $V_j \in S$ and $V_j$ is not a collider or (ii) $V_j$ is a collider and $V_j \notin S \wedge des(V_j, G) \notin S$. Two nodes are said to be *d*-separated by a conditioning set S if all the paths between the nodes are blocked by S. The *d*-separated nodes are independent of one another conditioned on set S[28].

*Causal structure learning (CSL):* Causal discovery is transformed to the problem of CSL[15] that concerns learning the adjacencies of nodes and the orientation of the edges in *G(V, E)* using an observational distribution. The idea of CSL is to utilize conditional independence in the data distribution to infer the structure of the causal graphical model. However, the same conditional independence relation can be satisfied by multiple causal models belonging to a Markov equivalence class. For example, the conditional independence relation $V_i \perp V_k / V_j$ is satisfied by the fork sub-graph $V_i \leftarrow V_j \rightarrow V_k$ as well as two chain sub-graphs $V_i \rightarrow V_j \rightarrow V_k$ and $V_i \leftarrow V_j \leftarrow V_k$. The causal structures entailing the same set of conditional independence relations belong to the Markov equivalence class. CSL generally concerns learning a Markov equivalence class of the underlying causal model. Additional parametric assumptions and domain knowledge are needed to identify a causal model within the Markov equivalence class.

*Assumptions:* In general, CSL methods make assumptions about the underlying data generating mechanism to learn a causal structure from observational data. Two common assumptions are: (1) *causal faithfulness* that implies that all the conditional independencies observed from the data distribution are entailed by the *d*-separation conditions of an underlying causal graph, and (2) *causal sufficiency* that refers to the absence of any unmeasured common causes of variables in *V*. Additionally, most CSL algorithms make assumptions of no selection bias and infinite sample size. These common assumptions may be relaxed by some algorithms. The next subsection briefly explicates the four causal discovery algorithms used in this study.

### 2.2. Causal Discovery algorithms

This subsection describes the four causal discovery algorithms used.

PC[30,31] is a constraint-based algorithm that uses conditional independencies in the data as constraints to estimate an equivalence class of the underlying SCM. It makes the causal sufficiency and causal faithfulness assumptions for the correctness of edge adjacencies. Then, *v*-structure discovery followed by Meek rules[32] produces an equivalence class graph, also known as a completed partial directed acyclic graph (CPDAG). An undirected edge $V_i - V_j$ in a CPDAG suggests both orientations $V_i \leftarrow V_j$ and $V_i \rightarrow V_j$ are possible for given data which can be oriented using the domain knowledge.

FCI[30] is another constraint-based algorithm that assumes causal faithfulness. FCI outputs a partial ancestral graph (PAG) to incorporate the hidden confounders using a bidirectional arrow, i.e., $V_i \leftrightarrow V_j$. In a PAG, $V_i \rightarrow V_j$ is interpreted as $V_i$ being an ancestor of $V_j$ and $V_j$ not being an ancestor of $V_i$. Similarly, $V_i \, o\!\!\rightarrow V_j$ indicates either $V_i$ is an ancestor of $V_j$ or there is a hidden confounder between $V_i$ and $V_j$.

FGES is an optimized and parallelized version of the Greedy Equivalence Search (GES)[33] algorithm. FGES is a score-based method that approaches CSL as the problem of fitting a causal graph that best describes the conditional independencies in the data using a relevant score function. GES starts with an empty graph and greedily keeps adding edges that increase the goodness-of-fit score. The algorithm then removes the edges until the score does not improve to return the equivalence class of DAGs with the maximum score. FGES makes the causal sufficiency assumption but allows some violation in the faithfulness assumption. FGES returns a CPDAG as output similar to PC.

DirectLiNGAM[34] is a variation of the causal discovery algorithm called Linear Non-Gaussian Acyclic Models (LiNGAM). LiNGAM[35] is a functional causal model-based (or equivalently structural equation model-based) CSL algorithm. The functional causal models[36] use SCM with additional assumptions on the distribution of U and V to distinguish between different DAGs in the same equivalence class. It assumes causal sufficiency, linear continuous data generating process, and exogeneous variables with non-Gaussian distributions of non-zero variance. The non-Gaussian nature of noise enables asymmetric cause-effect relationships that help in identification beyond the equivalence class.

Detailed explanation of the algorithms can be found in additional resources[15,36,37].

### 2.3. Structural Equation Model (SEM)

A SEM is a subclass of a structural causal model (SCM). The relationship between the target and explanatory variables are often represented as nonlinear and nonparametric functions in a SCM, whereas the functions in a SEM are often represented by linear relationships—this type of SEM is more precisely referred to as a *linear* SEM. In that sense, a SCM consists of a set of equations of the form $V_i = f_i(V_{pa}, U_i)$, where $f_i$ is a nonparametric, nonlinear generalization of the linear SEM $V_i = \alpha_i + BV_{pa} + U_i$, where $\alpha_i$ is vector of intercept terms for the equation and $B$ is the matrix of coefficient. Each equation in a SEM represents an autonomous mechanism, and if each variable has a distinct equation, then we can call the SEM an SCM[28].

A SEM has two components: a structural model and a measurement model. The former is related to the hypothetical assumptions about the relations between the latent variables, while the latter deals with connecting latent variables to observed variables[38]. Mathematically, the linear structural model can be represented as:

$$\eta_i = \alpha_n + B\eta_i + \Gamma\xi_i + \zeta_i \qquad (2)$$

where $\eta_i$ refers to vector of latent endogenous variable for unit $i$; $\alpha_n$ is vector of intercept terms for the equation; $B$ is the matrix of coefficient giving expected effects of latent endogenous variables ($\eta$) on each other; $\Gamma$ denotes the coefficient matrix giving the expected effects of the latent exogenous variable ($\xi$) on latent endogenous variables ($\eta$); $\zeta_i$ is vector of disturbances[38].

The measurement model can mathematically be represented by the following equations:

$$y_i = \alpha_y + \Lambda_y\eta_i + \varepsilon_i \qquad (3)$$

$$x_i = \alpha_x + \Lambda_x\xi_i + \delta_i \qquad (4)$$

where $y_i$ is the vector of observed indicator $\eta_i$; $x_i$ is the vector of observed indicator $\xi_i$; $\Lambda_y$ denotes matrix of factor loading or regression coefficients giving the impact of the latent variable $\eta_i$ on $y_i$; $\Lambda_x$ denotes matrix of factor loading or regression coefficients giving the impact of the latent variable $\xi_i$ on $x_i$; $\xi_i$ is the unique factors of $y_i$; $\delta_i$ is the unique factors of $x_i$[38].

In the literature, several metrics have been proposed to evaluate the performance of SEMs. These include Root Mean Square Error of Approximation (RMSEA), Comparative Fit Index (CFI), Goodness-of-Fit Index (GFI), Adjusted Goodness-of-Fit index (AGFI), Normed Fit Index (NFI),

Tucker-Lewis Index (TLI), Akaike information criterion (AIC), Bayesian Information Criterion (BIC), and Chi-Square ($\chi^2$) test. RMSEA is the indicator of the discrepancy of the model per degree of freedom[16]. CFI indicates the amount of variance accounted for in a covariance matrix[39]. GFI, AGFI, NFI, and TLI are all goodness-of-fit measures with variations in their formulations. AIC and BIC are the relative measures of the information lost when a model generates data and the extent of a model to be parsimonious respectively[39]. Chi-square test indicates the discrepancy in the model and hence it should preferably be non-significant[39].

In this study, we estimated the SEMs using polychoric correlation based Unweighted Least Square (ULS) estimation method which is the recommended method for ordered categorical data[40]. The reader is referred to additional resources[16,38–40] to know more about the estimation methods and goodness of fit measures.

### 2.4. Data

In this study, we used the 2017 National Household Travel Survey (NHTS) data[41]. These data are collected from a stratified random sample of U.S. households in all the 50 U.S. states and the District of Columbia. The data consists of information about each trip made by each household member on the household's travel day. Out of all the variables in the dataset, 14 variables were selected since these were suspected to affect the mode choice decision based on domain knowledge and previous studies[4,5]. These variables could be grouped into three categories, namely trip characteristics, trip attributes, and socio-demographic information. The variables were discretized and converted to binary and ordinal variables to fit the requirements of some of the causal discovery algorithms used in this study.

To narrow the scope of this study, only the trips that were made using cars, public transport, or walking were studied. Additionally, any respondents of age less than 18 years were removed. Further, any trip data with unknown values (for example, 'not ascertained', 'I don't know', 'I prefer not to answer', 'appropriate skip') were removed from the dataset. Since the availability of transportation infrastructure varies substantially throughout the country which could affect travel mode choices, we reduced the dataset to only the trip that occurred in New York Metropolitan area (also called the New York-Newark-Jersey City, NY-NJ-PA metropolitan statistical area). The cleaned dataset used in the analysis consisted of a total of 21,618 observations. Further, the data were scaled between 0 and 1 before applying the causal discovery algorithms. Table S1 in the supplementary materials presents the list of the variables, along with their description, coding, and percentage distribution.

### 2.5. Proposed methodology

In this study, we propose to use a combination of causal discovery and SEMs to model travel mode choice. This methodology involves the following steps:

*Step 1:* Survey data together with (obvious) domain knowledge is inputted into a causal discovery algorithm.

*Step 2:* Causal graphs are obtained as an output from the causal discovery algorithm.

*Step 3:* The causal graph and the survey data are fed into an SEM.

*Step 4:* The SEM estimates the direct causal effects between the variables.

*Step 5:* The performance of the SEM is judged by a set of goodness-of-fit measures.

*Step 6:* Steps 1-5 are repeated for a different causal discovery algorithm.

*Step 7:* The algorithm that provides the best performance compared to the rest of the algorithms in step 5 is selected as the final causal graph and the model results are interpreted.

Py-causal[42] and Lingam[43] python libraries were used for causal discovery algorithms, CausalNex[44] library was used to draw causal graphs, and the semopy[45] library was used for the SEMs. Figure 1 illustrates the proposed methodology.

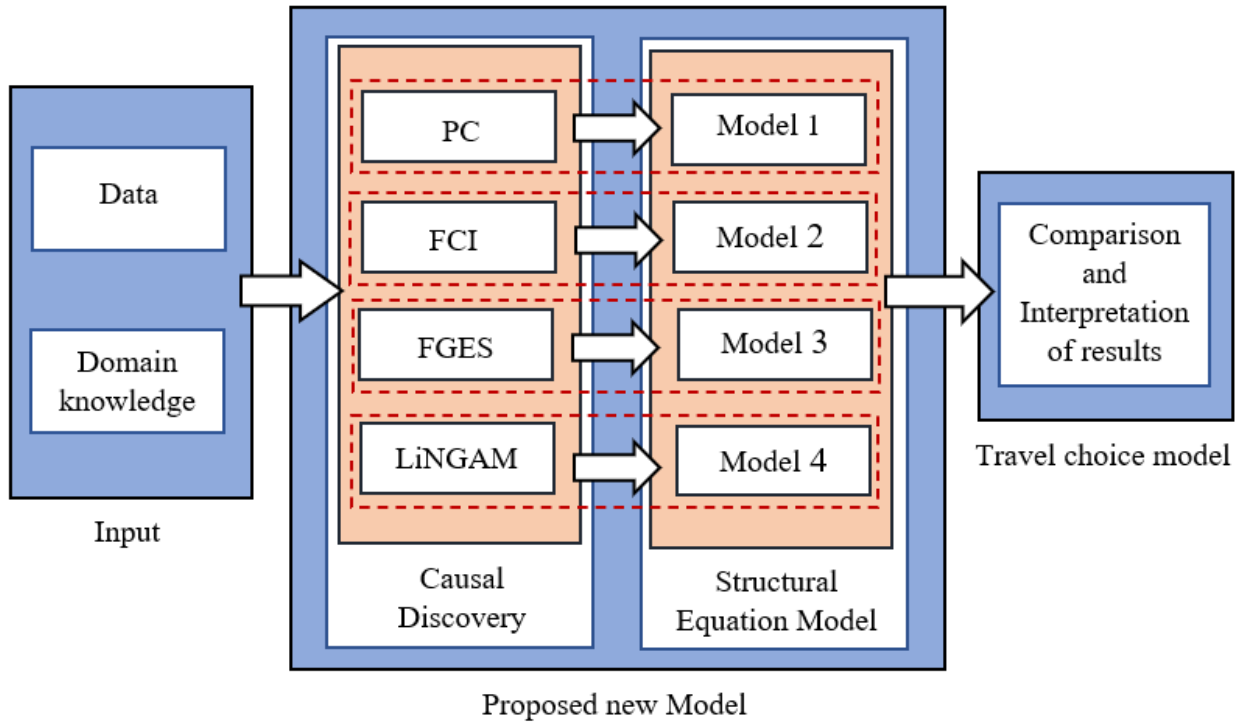

**Figure 1.** Diagrammatic representation of the proposed new methodology.

In step 1, domain knowledge was added to the causal discovery algorithms since they are known to improve the performance of the causal algorithms[5,13]. Therefore, some graphical restrictions were added to the causal graphs based on the domain expertise. Care was taken to minimize the number of such restrictions and to add only the ones that are obvious. The following domain knowledge restrictions were added:

- The three travel modes (car, public, and walk) were set as the target variable and hence were not allowed to cause any other variables.

- The trip characteristics were not allowed to cause the trip attributes and the socio-demographic variables.

- The trip attributes were assumed not to cause the socio-demographic variables.

- A few variables were assumed to be exogenous variables for the scope of this study. These were place type, gas price, race, age, and gender.

- Respondents' education level was assumed not to be caused by their worker status, vehicle ownership, and household income.

Since several causal discovery algorithms used in this study cannot detect latent confounders, for the sake of fair comparison, it was assumed that the causal mechanism behind mode choice decision making is not affected by any variables other than those mentioned in Table S1. In other words, we assume the causal Markov property and causal sufficiency for all the causal discovery algorithms.

## 3. Results

As a preliminary analysis, the correlations between the variables were computed. Since the variables are ordinal and binary, Spearman rank-order correlation was used. We found that the correlations were quite low, except for those between the travel modes. Only one correlation exceeds 0.5, between trip distance and walking (with a correlation value of -0.56). Thus, there were no highly correlated variables in the study data. Figure S1 in the supplementary materials shows the heat map of the correlations between the variables.

Several model evaluation metrics were used to compare the SEMs developed from each of the four causal models as shown in Table 1. The results show that DirectLiNGAM-based SEM outperforms other models based on the evaluation metrices. DirectLiNGAM also achieves accepted levels of CFI, GFI, AGFI, NFI, TLI, and RMSEA. None of the models obtained the preferred non-significant p-value for chi-square test. However, this could perhaps be due to the large size of the study dataset[39].

Based on the results from the evaluation metrices, it can be concluded that DirectLiNGAM-based SEM are the most reliable out of the other models tested in this study. Figure 2 shows the causal graph obtained from the DirectLiNGAM-based SEM model. The blue and red edges correspond to the positive and negative values of the corresponding path coefficients respectively. The thickness of the edges is made proportional to the magnitudes of the path coefficients. Due to space limitation, the graph is simplified by removing any edges with the value of path coefficient between 0.25 and -0.25. Figures S2-4 in the supplementary materials show the causal graphs obtained for the PC-based, FCI-based, and FGES-based SEM models. Tables S2-4 show the model output from each of these models. Figure S5 and Table S5 present the complete graphical causal model and model results from the DirectLiNGAM-based SEM which is found to be most accurate.

Given that the DirectLiNGAM-based SEM model was built on causal structural graph and has passed the goodness-of-fit tests, the interpretation of the result from this model can be done causally. Noted computer scientist Judea Pearl has advocated for interpreting SEMs causally[20]. On SEMs, he explained that "The 'path coefficient,' $\beta$, quantifies the (direct) causal effect of X on Y. Once we commit to a particular numerical value of $\beta$, the equation claims that a unit increase for X would result in $\beta$ units increase of Y regardless of the values taken by other variables in the model, regardless of the statistics of $U_X$ and $U_Y$, and regardless of whether the increase in X originates from external manipulations or variations in $U_X$."[20] Here $U_X$ and $U_Y$ denotes exogenous variables in the model.

**Table 1.** Modeling performance of the various SEMs. Bolded figures show the best result.

| | PC-based | FCI-based | FGES-based | Direct LiNGAM-based | Standard accepted levels |
|---|---|---|---|---|---|
| Comparative fit index (CFI) | 0.922 | 0.803 | 0.942 | **0.986** | $\geq 0.95$ |
| Goodness-of-fit index (GFI) | 0.922 | 0.803 | 0.941 | **0.986** | Closer to 1.0 is preferred. |
| Adjusted goodness-of-fit index (AGFI) | 0.896 | 0.755 | 0.918 | **0.960** | Closer to 1.0 is preferred. |
| Normed fit index (NFI) | **0.922** | 0.803 | **0.941** | 0.986 | >0.90 |
| Tucker-Lewis index (TLI) | 0.897 | 0.755 | **0.918** | 0.960 | >0.90 |
| Root Mean Square Error of Approximation (RMSEA) | 0.094 | 0.155 | 0.084 | **0.055** | <0.06 |
| Akaike information criterion (AIC) | 71.99 | 10.10 | 84.26 | **-3.15· $10^{16}$** | Lower the better |
| Bayesian information criterion (BIC) | 399.22 | 209.64 | 467.36 | **-3.15· $10^{16}$** | Lower the better |
| Log likelihood | 5.01 | 19.95 | 5.87 | 1.57· $10^{16}$ | |
| Degree of freedom | 95 | 66 | 88 | 50 | |
| Degree of freedom baseline | 126 | 82 | 124 | 141 | |
| Chi-squared test | 18111 | 34326 | 13408 | 3326 | |
| p-value for chi-squared test | 0.000 | 0.000 | 0.000 | 0.000 | >0.05 |
| Chi-squared test baseline | 231024 | 174027 | 229090 | 231594 | |
| Number of links/edges | 31 | 16 | 36 | 91 | |

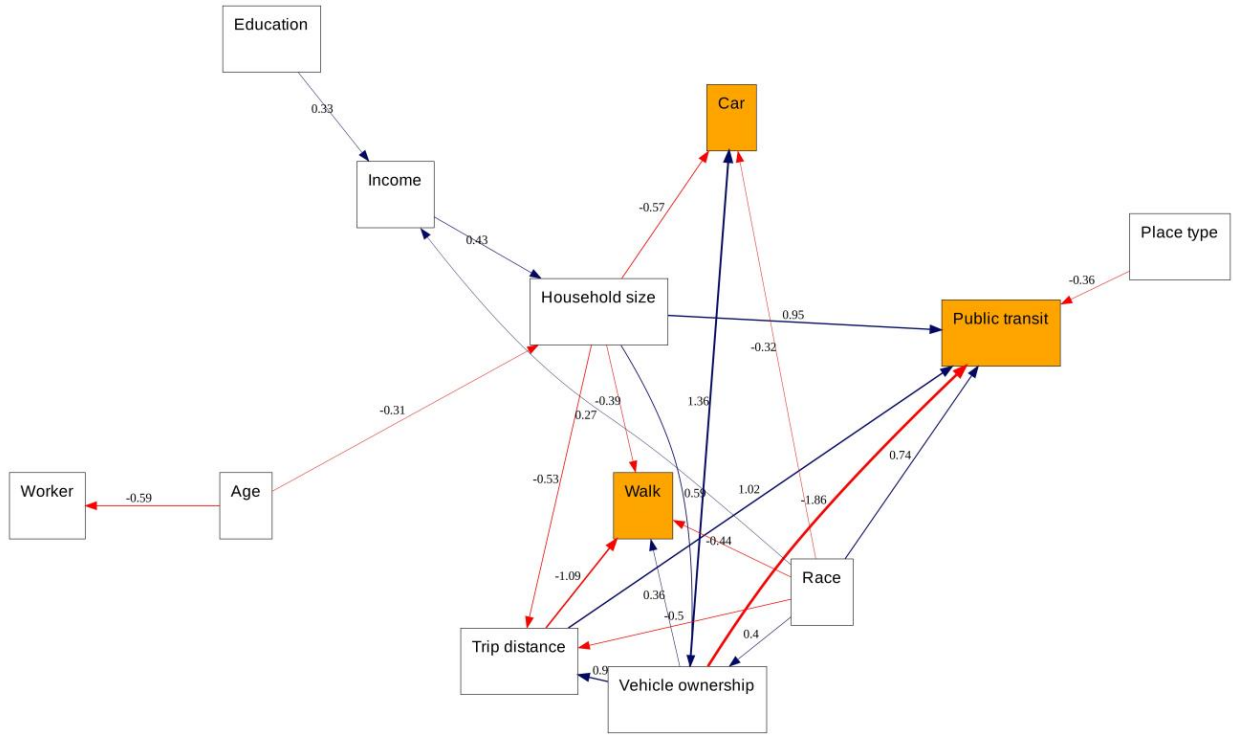

**Figure 2.** Simplified results from the DirectLiNGAM-based SEM.

The rest of this section discusses the results from the selected DirectLiNGAM-based SEM, focusing on mode choice decision making.

Based on the values of path coefficients, choosing car as a travel mode has strong positive direct causal effect from number of vehicles owned (1.357) and trip distance (0.181). It has strong negative causal effects from household size (-0.573), white race (-0.319), and household income (-0.215). Further, small direct causal effects (path coefficients between 0.1 and -0.1) are caused by place type, education level, peak hour, age, gender, home-based trip, worker status, and weekday.

Similarly, choosing public transit as travel mode has strong positive direct causal effect from trip distance (1.018), household size (0.951), white race (0.742), and weekday (0.123). It has strong negative causal effect from number of vehicles owned (-1.858), place type (-0.356), and home-based trip (-0.103). Small direct causal effects are caused by household income, gas price, peak hour, age, and worker status.

Lastly, choosing walking has strong positive direct causal effect from number of vehicles owned (0.361), place type (0.243), and home-based trip (1.72). It has strong negative causal effect from trip distance (-1.093), white race (-0.436), and household size (-0.387). Small direct causal effects are caused by household income, worker status, gender, education, weekday, gas place, and age. These results seem to be logically correct.

## 4. Discussion

Over recent years, the field of transportation has seen a growing popularity in using advanced techniques from statistics, machine learning, and deep learning[46–50]. Almost all these techniques are based on correlations. Introducing the concept of causality in transportation is a step forward and to do so, this study introduces a novel methodology for causal modeling of travel modes. Several noteworthy observations were made during this analysis which highlights the challenges and opportunities in the application of causality in transportation modeling. These include:

i. **Comparison between the causal discovery algorithms:** The four causal discovery algorithms tested in this study differ in their assumptions and methodology. These differences are reflected in the causal graphs they produced. These causal graphs varied drastically in terms of their complexity. DirectLiNGAM produced the most complex causal graph while FCI generated the simplest of all. The number of edges in the causal graph from DirectLiNGAM are more than five times larger than that from FCI. The algorithms agreed on some while disagreed on other cause-and-effect pairs. For example, all four algorithms agreed that number of vehicles owned and trip distance have a direct causal effect on choosing public transit as well as on choosing to walk. In contrast, no single common variable was found to be a direct cause of choosing car from all four algorithms. All the algorithms were found to benefit from the domain knowledge that was inputted to them.

ii. **Further potential in FCI:** The FCI algorithm performed the worst in almost all the evaluation metrices. Nevertheless, this might be because the full potential of FCI was not explored in this study. FCI does not hold the causal sufficiency assumption and hence can suggest the presence of unobserved cofounders. In our analysis, FCI discovered three unobserved latent variables (Figure S3) and indicated the possibility of some more. For comparison purposes, we assumed that the causal mechanism behind mode choice decision making is not affected by any other variables than the 14 variables selected in this study (Table S1). This limited the usefulness of FCI. Future studies can explore the potential of FCI in detecting unobserved latent variables in transportation modeling.

iii. **Uncovering complex societal issues using causal discovery:** Unless restricted, the causal discovery algorithms find causal relations between all the variables in the dataset. This could lead to the detection of some nontrivial causal connections. For instance, all four algorithms found a causal link from race to number of vehicles owned, and from race to household size. This highlights the potential of causal discovery in discovering complex socio-economic issues. While the analysis of the validity of these causal links is beyond the scope of this study, we do not suggest that race is a cause of any of the travel behavior associated variables. The variable 'race' was kept in this study because of a long-established practice of including race in mode choice modeling, which is expected to serve as a proxy for a complex mix of socio-cultural, political, and economical factors, instead of racial/ethnical differences. Further, it was found that removing race from the analysis decreases the performance of the models substantially. The causal effects from variable 'race' also emphasizes on the care that must be taken while interpreting causal discovery models which might be based on the assumption of causal sufficiency. Causal sufficiency implies that there exist no unobserved confounders between the variables. However, it is highly likely that there are some unobserved exogenous variables that exist between race and other variables.

iv. **Presence of non-manipulable variables:** Some researchers believe in 'no causation without manipulation' and are skeptical about determining causal effects from non-manipulable variables like gender and race[22,51]. Other researchers have found this practice legitimate[22,51]. However, based on the arguments proposed by Pearl[51], we decided to keep non-manipulable variables in our models. Pearl[51] suggested that even if some variables may be non-manipulable, knowing their causal effects can be useful for guiding policy making.

v. **Caveats of the proposed methodology:** In this study, we generated causal graphs from causal discovery algorithms and (minimum) domain knowledge. These causal graphs were used to estimate several SEMs. The SEM that fits the data best was selected as the final graph (which in our analysis was the DirectLiNGAM-based SEM). Our methodology makes an important contribution in laying out the procedure for causal modeling. However, it must be noted that the final graph selected may not be the ultimate perfect causal graph and there is always the possibility to find a more accurate graph. Bollen and Pearl[22] explained this property of SEM as "Fitting the data does not 'prove' the causal assumptions, but it makes them tentatively more plausible. Any such positive results need to be replicated and to withstand the criticisms of researchers who suggest other models for the same data."[22] Despite this caveat, we argue that the same is true with any other transportation data modeling technique. Next, it must also be noted that causal discovery algorithms and SEMs are based on certain assumptions and limitations. One of these is that the algorithms used in this study cannot detect bidirectional cause-and-effect links; i.e., the causal graph do not have a provision of feedback loops. This could be an interesting research topic for future studies. Another important assumption that we made was that there are no variables other than those selected by the authors (Table S1) that affects mode-choice decision. However, it is likely that there can be more variables in the final causal graph (simplified in Figure 2 and full graph in Figure S5) that were out of the scope of this study.

vi. **Contribution of the proposed methodology:** The long-established approach for mode choice modeling is correlation based. SEM modeling has been proposed as an alternative approach[16–19]. SEMs could be considered a causal model; however, its accuracy depends on the accuracy of the hypothesized causal graph supplied to them[22]. These hypothesized causal graphs have almost always been deduced from domain knowledge. Our proposed methodology focuses on causal modeling by advancing the SEM approach by combining it with causal discovery algorithms. This combination is mutually beneficial. Causal discovery algorithms can extract causal graphs directly from the observational data, which can be used as an input to the SEM. Subsequently, the SEM can estimate the quantitative direct causal effects for the causal graph. In addition, the goodness of fit measures obtained from the SEM models can be used to conduct a comparative evaluation of the performance of the various causal discovery algorithms. To do such a comparison, previous studies had to rely on the proximity of a causal graph with the 'true' causal graph. This true causal graph is created based on the expert's knowledge or from the knowledge of data creating process (e.g., in simulation studies)[13,15]. Yet, creating a true causal graph is often challenging, if not impossible, in complex real-world scenarios. Our suggested methodology, therefore, provides a data-driven approach to compare the performance of the various causal discovery algorithms. Further, this methodology advances mode choice modeling and is a step forward towards developing reliable causal models as opposed to the traditional correlation-based approach. The resulting causal graph presents the visual representation of the complexity involved in mode choice decision making. The variables known to be affecting mode choice were found to have mutual causal connections. This contradicts the assumption made by most of the statistical and machine learning approaches to mode choice modeling. Our modeling results suggest that the novel methodology is intensive and dependable.

## 5. Conclusion

Overall, the causal discovery-based SEM is a dependable methodology. The biggest advantage of this approach is that it helps SEMs by providing them with data-driven causal graphs making them more objective. Our study found that DirectLiNGAM is the most accurate algorithm for mode choice modeling out of the four algorithms tested. In the comparative analysis, the DirectLiNGAM-based SEM passes several goodness-of-fit tests like RMSEA, CFI, GFI, AGFI, NFI, and TLI, and achieves the lowest AIC and BIC values. The implementation of this DirectLiNGAM-based SEM method provided insights to the complex processes involved in travel decision making. The study identified several logically reasonable variables that cause travel mode choice.

The study has potential for improvement. The most notable one could be to include latent or unobserved confounding variables. Further, the level-of-service attributes of the different modes were not included in the model due to data limitations. Finally, testing the proposed framework for other mobility choice contexts, for instance residential location, can help to generalize the findings. These improvements can be part of future research.

## Author Contributions

The authors confirm contribution to the study as follows: study conception and design: RSC, SD, EZ, CFC, and FCP; Analysis: RSC, CR, and SA; Draft manuscript preparation: RSC, CR, and SA. All authors interpreted/reviewed the results and approved the final version of the manuscript.

## Acknowledgement

This work was supported in part by the United States National Science Foundation (NSF) under Grant No. 1551731.

## Data Availability

The dataset analysed during the current study were conducted on NHTS data[41] which are available at https://nhts.ornl.gov/.

## Competing interests

The authors do not have any conflict of interest.

# References

1. Derrible, S. *Urban engineering for sustainability*. (MIT Press, 2019).
2. Zhao, X., Yan, X., Yu, A. & Van Hentenryck, P. Prediction and behavioral analysis of travel mode choice: A comparison of machine learning and logit models. *Travel Behav. Soc.* **20**, 22–35 (2020).
3. Cheng, L., Chen, X., De Vos, J., Lai, X. & Witlox, F. Applying a random forest method approach to model travel mode choice behavior. *Travel Behav. Soc.* **14**, 1–10 (2019).
4. Xie, C. & Waller, S. Estimation and application of a Bayesian network model for discrete travel choice analysis. *Transp. Lett.* **2**, 125–144 (2010).
5. Ma, T.-Y., Chow, J. Y. & Xu, J. Causal structure learning for travel mode choice using structural restrictions and model averaging algorithm. *Transp. Transp. Sci.* **13**, 299–325 (2017).
6. Xie, C., Lu, J. & Parkany, E. Work travel mode choice modeling with data mining: decision trees and neural networks. *Transp. Res. Rec.* **1854**, 50–61 (2003).
7. Lee, D., Mulrow, J., Haboucha, C. J., Derrible, S. & Shiftan, Y. Attitudes on autonomous vehicle adoption using interpretable gradient boosting machine. *Transp. Res. Rec.* **2673**, 865–878 (2019).
8. Lee, D., Derrible, S. & Pereira, F. C. Comparison of four types of artificial neural network and a multinomial logit model for travel mode choice modeling. *Transp. Res. Rec.* **2672**, 101–112 (2018).
9. Pearl, J. & Mackenzie, D. *The book of why: the new science of cause and effect*. (Basic books, 2018).
10. Hujoel, P., Cunha-Cruz, J. & Kressin, N. Spurious associations in oral epidemiological research: the case of dental flossing and obesity. *J. Clin. Periodontol.* **33**, 520–523 (2006).
11. Listl, S. & Chiavegatto Filho, A. D. Big data and machine learning. in *Oral Epidemiology* 357–365 (Springer, 2021).
12. Vigen, T. *Spurious correlations*. (Hachette UK, 2015).
13. Shen, X., Ma, S., Vemuri, P. & Simon, G. Challenges and opportunities with causal discovery algorithms: application to Alzheimer's pathophysiology. *Sci. Rep.* **10**, 1–12 (2020).
14. Nogueira, A. R., Pugnana, A., Ruggieri, S., Pedreschi, D. & Gama, J. Methods and tools for causal discovery and causal inference. *Wiley Interdiscip. Rev. Data Min. Knowl. Discov.* **12**, e1449 (2022).
15. Heinze-Deml, C., Maathuis, M. H. & Meinshausen, N. Causal Structure Learning. *Annu. Rev. Stat. Its Appl.* **5**, 371–391 (2018).
16. Wang, Y., Yan, X., Zhou, Y. & Xue, Q. Influencing mechanism of potential factors on passengers' long-distance travel mode choices based on structural equation modeling. *Sustainability* **9**, 1943 (2017).
17. Golob, T. F., Kitamura, R. & Supernak, J. A panel-based evaluation of the San Diego I-15 Carpool Lanes Project. in *Panels for Transportation Planning* 97–128 (Springer, 1997).
18. Golob, T. F. & Hensher, D. A. Greenhouse gas emissions and Australian commuters' attitudes and behavior concerning abatement policies and personal involvement. *Transp. Res. Part Transp. Environ.* **3**, 1–18 (1998).
19. Levine, J., Park, S., Wallace, R. R. & Underwood, S. E. Public choice in transit organization and finance: The structure of support. *Transp. Res. Rec.* **1669**, 87–95 (1999).
20. Pearl, J. *The causal foundations of structural equation modeling*. (2012).

21. Golob, T. F. Structural equation modeling for travel behavior research. *Transp. Res. Part B Methodol.* **37**, 1–25 (2003).
22. Bollen, K. A. & Pearl, J. Eight myths about causality and structural equation models. in *Handbook of causal analysis for social research* 301–328 (Springer, 2013).
23. Tarka, P. An overview of structural equation modeling: its beginnings, historical development, usefulness and controversies in the social sciences. *Qual. Quant.* **52**, 313–354 (2018).
24. McCoach, D. B., Black, A. C. & O'Connell, A. A. Errors of inference in structural equation modeling. *Psychol. Sch.* **44**, 461–470 (2007).
25. Kelloway, E. K. Structural equation modelling in perspective. *J. Organ. Behav.* **16**, 215–224 (1995).
26. MacCallum, R. C., Roznowski, M. & Necowitz, L. B. Model modifications in covariance structure analysis: the problem of capitalization on chance. *Psychol. Bull.* **111**, 490 (1992).
27. Monteiro, M. M. Adaptation of transnational short-term residents: understanding the factors influencing residential location choice and travel behavior. (2020).
28. Pearl, J. *Causality*. (Cambridge university press, 2009).
29. Pearl, J. *Probabilistic reasoning in intelligent systems: networks of plausible inference*. (Morgan kaufmann, 1988).
30. Sprites, P., Glymour, C. & Scheines, R. Causation, prediction and search (First, online ed.). (1993).
31. Colombo, D. & Maathuis, M. H. Order-independent constraint-based causal structure learning. *J Mach Learn Res* **15**, 3741–3782 (2014).
32. Meek, C. Causal inference and causal explanation with background knowledge In Uncertainty in Artificial Intelligence 11. (1995).
33. Chickering, D. M. Optimal structure identification with greedy search. *J. Mach. Learn. Res.* **3**, 507–554 (2002).
34. Shimizu, S. *et al.* DirectLiNGAM: A direct method for learning a linear non-Gaussian structural equation model. *J. Mach. Learn. Res.-JMLR* **12**, 1225–1248 (2011).
35. Shimizu, S., Hoyer, P. O., Hyvärinen, A., Kerminen, A. & Jordan, M. A linear non-Gaussian acyclic model for causal discovery. *J. Mach. Learn. Res.* **7**, (2006).
36. Glymour, C., Zhang, K. & Spirtes, P. Review of causal discovery methods based on graphical models. *Front. Genet.* **10**, 524 (2019).
37. Tetrad Single HTML Manual. https://cmu-phil.github.io/tetrad/manual/#alpha.
38. Bollen, K. A. & Noble, M. D. Structural equation models and the quantification of behavior. *Proc. Natl. Acad. Sci.* **108**, 15639–15646 (2011).
39. Fan, Y. *et al.* Applications of structural equation modeling (SEM) in ecological studies: an updated review. *Ecol. Process.* **5**, 1–12 (2016).
40. Xia, Y. & Yang, Y. RMSEA, CFI, and TLI in structural equation modeling with ordered categorical data: The story they tell depends on the estimation methods. *Behav. Res. Methods* **51**, 409–428 (2019).
41. National Household Travel Survey. https://nhts.ornl.gov/.
42. Wongchokprasitti, C. (Kong) *et al.* bd2kccd/py-causal v1.2.0 | Zenodo. https://zenodo.org/record/3364590/export/dcat#.YuNaCD3MJPZ.
43. LiNGAM - Discovery of non-gaussian linear causal models. (2022).
44. Welcome to CausalNex's API docs and tutorials! — causalnex 0.11.0 documentation. https://causalnex.readthedocs.io/en/latest/.

45. Igolkina, A. A. & Meshcheryakov, G. semopy: A Python Package for Structural Equation Modeling. *Struct. Equ. Model. Multidiscip. J.* **27**, 952–963 (2020).
46. Chauhan, R. S. Short-Term Traffic Delay Prediction at the Niagara Frontier Border Crossings Using Deep Learning. (2019).
47. Chauhan, R., Shi, Y., Bartlett, A. & Sadek, A. W. Short-Term Traffic Delay Prediction at the Niagara Frontier Border Crossings: Comparing Deep Learning and Statistical Modeling Approaches. *J. Big Data Anal. Transp.* **2**, 93–114 (2020).
48. Taghipour, H., Parsa, A. B., Chauhan, R. S., Derrible, S. & Mohammadian, A. K. A novel deep ensemble based approach to detect crashes using sequential traffic data. *IATSS Res.* **46**, 122–129 (2022).
49. Lin, L. *et al.* Developing Predictive Border Crossing Delay Models. (2019).
50. Parsa, A. B., Movahedi, A., Taghipour, H., Derrible, S. & Mohammadian, A. K. Toward safer highways, application of XGBoost and SHAP for real-time accident detection and feature analysis. *Accid. Anal. Prev.* **136**, 105405 (2020).
51. Pearl, J. Does obesity shorten life? Or is it the soda? On non-manipulable causes. *J. Causal Inference* **6**, (2018).